\documentclass{aastex62}
\usepackage{amsmath}
\usepackage{color}
\usepackage{natbib}
\usepackage{amssymb}
\usepackage{amsfonts}
\usepackage{stackengine}
\usepackage{changepage}
\usepackage{subfigure}
\usepackage{graphicx}
\usepackage{rotating}

\newcommand{\beq}{\begin{equation}}
\newcommand{\eeq}{\end{equation}}

\def\ba{\begin{eqnarray}}
\def\ea{\end{eqnarray}}
\newcommand{\tg}{T_{\rm GRB}}
\newcommand{\tp}{T_{\rm pre}}
\newcommand{\tw}{T_{\rm wt}}
\newcommand{\db}{\Delta {\rm BIC}}

\shorttitle{Precusor of Short GRBs}
\shortauthors{Wang et al.}
\begin{document} 

\title{Stringent Search for Precursor Emission in Short GRBs from {\em Fermi}/GBM data and Physical Implications}

\author[0000-0002-2662-6912]{Jie-Shuang Wang}
\affiliation{Tsung-Dao Lee Institute, Shanghai Jiao Tong University, Shanghai 200240, China}
\author{Zong-Kai Peng}
\affiliation{School of Astronomy and Space Science, Nanjing
	University, Nanjing 210093, China}
\affiliation{Key Laboratory of Modern Astronomy and Astrophysics (Nanjing University), Ministry of Education, China}
\author{Jin-Hang Zou}
\affiliation{School of Astronomy and Space Science, Nanjing
	University, Nanjing 210093, China}
\affiliation{Key Laboratory of Modern Astronomy and Astrophysics (Nanjing University), Ministry of Education, China}
\author[0000-0003-4111-5958]{Bin-Bin Zhang}
\affiliation{School of Astronomy and Space Science, Nanjing
University, Nanjing 210093, China}
\affiliation{Key Laboratory of Modern Astronomy and Astrophysics (Nanjing University), Ministry of Education, China}
\affiliation{Department of Physics and Astronomy, University of Nevada Las Vegas, NV 89154, USA}
\author[0000-0002-9725-2524]{Bing Zhang}
\affiliation{Department of Physics and Astronomy, University of Nevada Las Vegas, NV 89154, USA}
\correspondingauthor{Jie-Shuang Wang; Bin-Bin Zhang}
\email{jiesh.wang@gmail.com; bbzhang@nju.edu.cn}

\begin{abstract}

We perform a stringent search for precursor emission of short gamma-ray bursts (SGRBs) from the {\em Fermi}/GBM data and find 16 precursor events with  $\gtrsim4.5\sigma$ significance. 
We find that the durations of the main SGRB emission ($\tg$) and the precursor emission ($\tp$), as well as the waiting time ($\tw$) in between, are roughly comparable to each other, with $\tw\approx2.8\tg^{1.2}$  approximately satisfied for most cases except one significant outlier. 
We also perform spectral analyses to the precursors and SGRBs, and find that the spectra of precursor emission can be fitted with the blackbody, non-thermal cutoff power law and/or power law models. We consider several possible models for precursor emission in SGRBs and find that the luminosity and spectral shape may be explained by the the shock breakout or the photospheric radiation of a fireball launched after the merger for thermal precursors, or magnetospheric interaction between two NSs prior to the merger for non-thermal precursors. For the fireball photospheric model, a matter-dominated jet is preferred and a constraint on the fireball Lorentz factor can be placed as $\Gamma\sim30$. For the magnetospheric interaction model, jet launching mechanism may be constrained. In particular, those events with $\tw/\tg\gg1$ (e.g. GRB191221802) require the formation of a supramassive or stable neutron star after the merger, with the delay time defined by the timescale for an initially baryon-loaded jet to become magnetically dominated and relativistic.

\end{abstract}

\section{Introduction}
The joint detection of a gravitational wave (GW) event GW170817 and a short gamma-ray burst (SGRB) GRB 170817A confirms that at least some SGRBs originate from double neutron star (NS) mergers \citep{Abbott2017a,Abbott2017b,Goldstein2017,Zhang2018}. 
Later, another NS merger event GW190425 was discovered  \citep{Abbott2020}, and a sub-threshold GRB, GBM-190816, was reported to be possibly associated with a sub-threshold GW event \citep{Yang2019,Goldstein2019}. 
While the GW observations alone can provide constraints on the NS equation of state \citep[e.g.][]{Abbott2017a,Abbott2020}, the joint GW-EM detections would provide further useful information about the physics of NS mergers and associated GRBs, including the jet launching mechanism, jet structure, jet composition, as well as GRB radiation mechanism \cite[e.g.][]{Troja2017,Zhang2018,Mooley2018a,Mooley2018b,Gill2019,Geng2019,Zhang2019,Yang2019,Troja2019,Ryan2020}.

Copious electromagnetic (EM) signals are expected to be generated before and after the NS merger \citep[for reviews, see][]{Berger2014,Fernandez2016,Zhang2018book,Metzger2019}. 
Prior to the merger, EM signals can be produced by the interaction between the magnetospheres of the two NSs \citep{Hansen2001,Lai2012,Palenzuela2013,Wang2016,Wang2018} or possible crust cracking of one or both NSs \citep{Tsang2012,Suvorov2020}. 
These mechanisms can lead to gamma-ray signals, which could be observed as precursor emission of SGRBs \citep[e.g.][]{Troja2010,Wang2018}. 
Precursor emission of SGRBs can also be produced after the merger. If the main SGRB emission is produced by the standard GRB mechanism (e.g. synchrotron radiation in an internal shock or magnetic dissipation site), a thermal precursor may be produced either as the shock breaks out from the surrounding ejecta or as the fireball ejecta reaches its photosphere radius \citep[e.g.][]{Meszaros2000,Ramirez-Ruiz2002}.

Many efforts have been made to search for precursor emission for GRBs. Precursors were firstly identified in long GRBs \citep[e.g.][]{Lazzati2005,Hu2014,Zhang2018a}. 
For NS-merger-origin SGRBs, intense, short $\gamma$-ray emission is expected to occur shortly after the merger. The detection of precursor emission is therefore of great interest to diagnose the physical process right before or shortly after the merger. Observationally, identifying a weak signal before the main SGRB signal often suffers from instrumental biases, such as the energy range and sensitivity. {\em Fermi}/Gamma-ray Burst Monitor (GBM) covers a broad energy band (from $\sim8$ keV to $40$ MeV), while {\em Swift}/Burst Alert Telescope (BAT) is more sensitive in the $15-150$ keV range. Thus {\em Swift}/BAT would have a higher rate to detect soft weak precursors. Indeed, \cite{Troja2010} found that $\sim8\%-10\%$ SGRBs detected by {\em Swift}/BAT are associated with precursor activities, 
while in the SPI-ACS/INTEGRAL SGRB catalog, only $<0.4\%$ of the SGRBs are found to have precursor emission \citep{Minaev2017}\footnote{One event, GRB 100717, was actually regarded as a long GRB in the {\em Fermi}/GBM catalog. However, \cite{Wang2018} analyzed the spectra of GRB 100717 and found that this event can be well explained as an SGRB with a precursor generated by magnetospheric interaction between two merging NSs.}.
Recently, \cite{Zhong2019} analyzed the {\em Swift} and {\em Fermi}/GBM SGRB data and found that $2.7\%$ of SGRBs have precursor emission. 

In this paper, we study the precursor emission of SGRBs in detail both observationally and theoretically. In Section 2, we first perform a systematic search for precursors in the {\em Fermi}/GBM SGRB catalog and then perform detailed data analyses to extract the temporal and spectral information of both the precursor and the main SGRB emission. 
In Section 3, we discuss the validity of several precursor models and constrain these models using observations. The conclusion and discussion are presented in Section 4.

\section{Data analysis and results}
\subsection{Precursor emission in {\em Fermi}/GBM SGRB sample}

SGRBs are usually classified based on the duration criterion $T_{90}\lesssim 2$ s. However, since the duration of GRB 170817A (associated with GW170817) is  $2.05$ s \citep{Zhang2018}, in this paper we adopt a more conservative criterion $T_{90}\lesssim3$ s to identify SGRB candidates. Up to April 2020, {\em Fermi}/GBM detected 529 such SGRB candidates \citep[][see also the online catalog\footnote{https://heasarc.gsfc.nasa.gov/W3Browse/fermi/fermigbrst.html}]{Bhat2016}.
GBM consists of twelve sodium iodide (NaI) detectors (sensitive to 8 keV - MeV band) pointing to different directions and two bismuth germanate (BGO) detectors (sensitive to 200 keV - 40 MeV band). 
We sort out two NaI detectors that have the smallest angular separations with respect to the sky position of the corresponding GRB. 
The Time-Tagged Event data from the two NaI detectors are used to construct the light curve, which provides the arrival times and photon energies. 
We select the data with photon arrival time between $T_0-50$ s and $T_0+30$ s, where $T_0$ is the GRB trigger time. 
Using the Bayesian Block (BB) algorithm \citep{Scargle2013} in the Astropy package \citep{Astropy2018}, we segment the photons into a sequence of time blocks, as shown in Fig. \ref{fig:1}. 
We then search for precursor emission in these time block sequences.

A precursor is defined as the first pulse in the light curve. It must satisfy the following three requirements: (1) the peak flux is lower than that of the main pulse; (2) the flux during the waiting time period (the time interval between the precursor and the main pulse) is consistent with the background level; 
(3) the significance level is larger than $3\sigma$. 
The first two requirements are the common definitions to identify precursor emission in SGRBs. 
The last one is reinforced in our study to reduce false-alarm signals. 
The second or main pulse is regarded as the main SGRB. 
To further strengthen the connection between the SGRB and the precursor, we also examined whether the precursor emission is only significant in the detectors in which the main pulse is bright. 
Then we follow the common definition of $T_{90}$ to calculate the durations of the precursor emission $(\tp)$ and the main SGRB ($\tg$), as well as the waiting time $(\tw)$ in between.
The significance level of the precursor depends on time-bin size, energy band, and the background level. We take the background data from two time intervals, i.e., 30 s before the precursor and 30 s after the main SGRB. We then simultaneously vary the energy band and bin size (limited to $<0.5\tp$) to determine the maximum significance level.

\subsection{Properties of the precursor and the main SGRB emission}
\begin{figure}[tbp]
\centering
	\includegraphics[angle=0,width=0.3\textwidth]{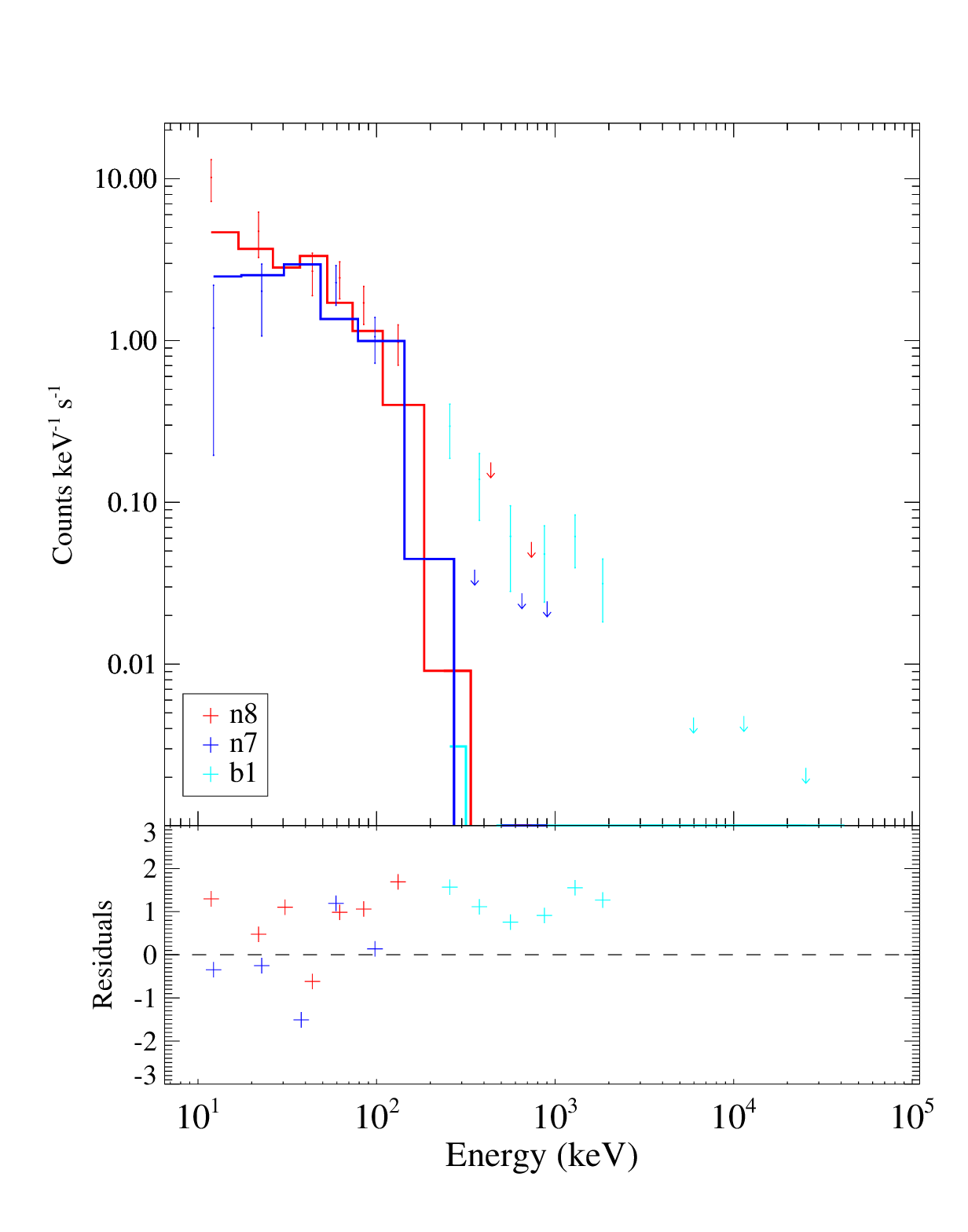}
	\includegraphics[angle=0,width=0.3\textwidth]{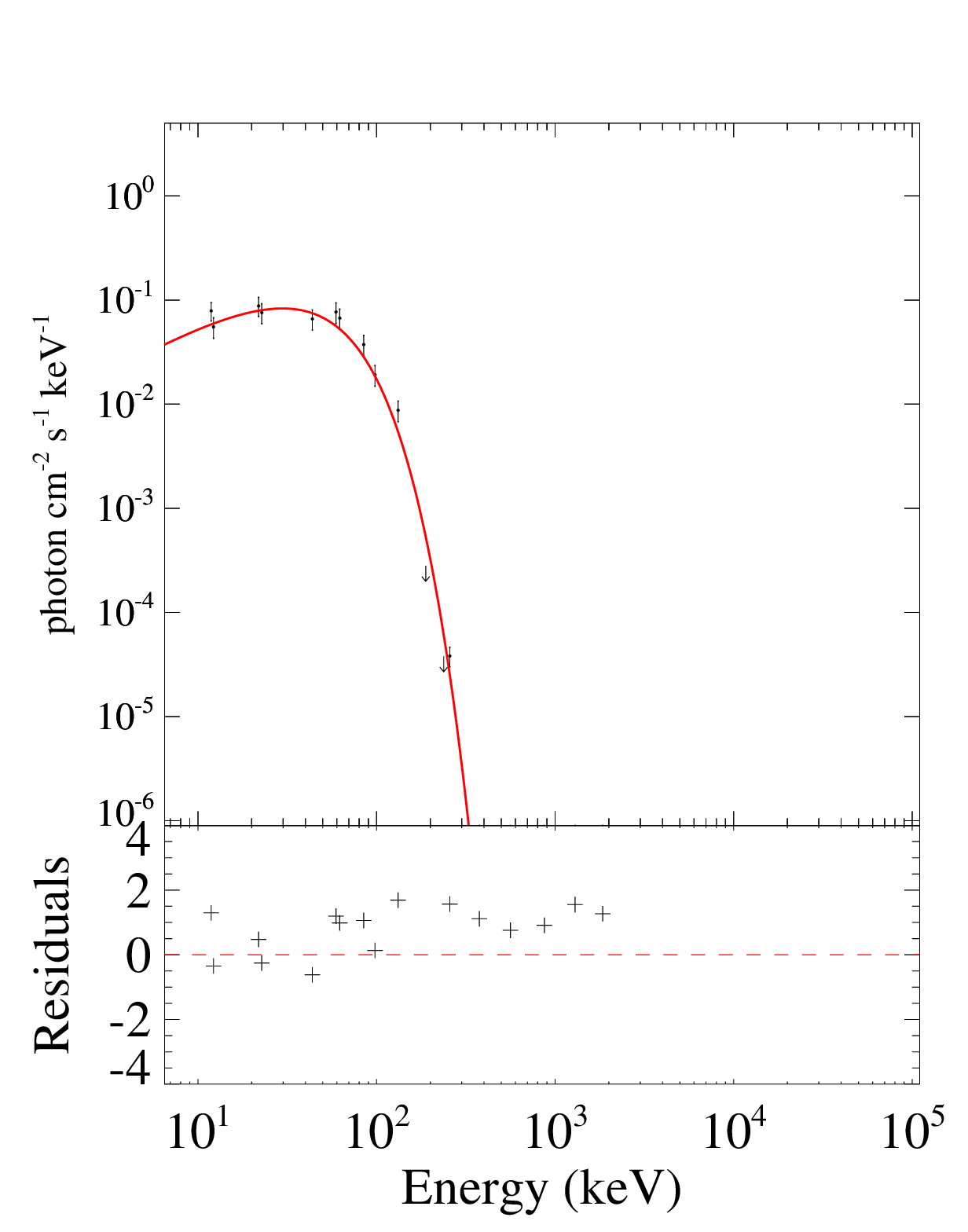}
	\includegraphics[angle=0,width=0.38\textwidth]{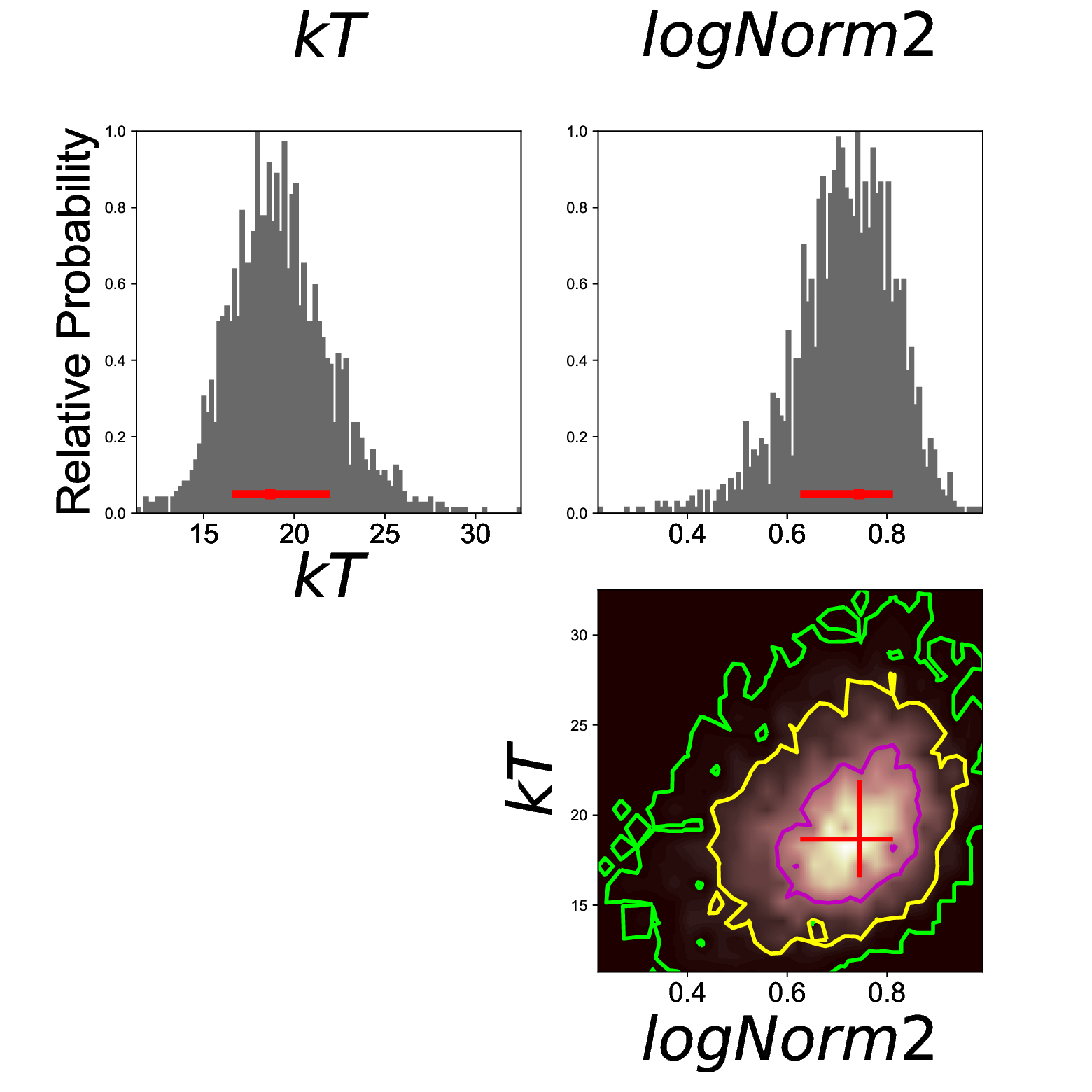}
	\caption{Constraints on the blackbody model of the precursor of GRB081216531.}\label{fig:GRB081216531}
\end{figure}

\begin{figure}[tbp]
\centering
	\includegraphics[angle=0,width=0.3\textwidth]{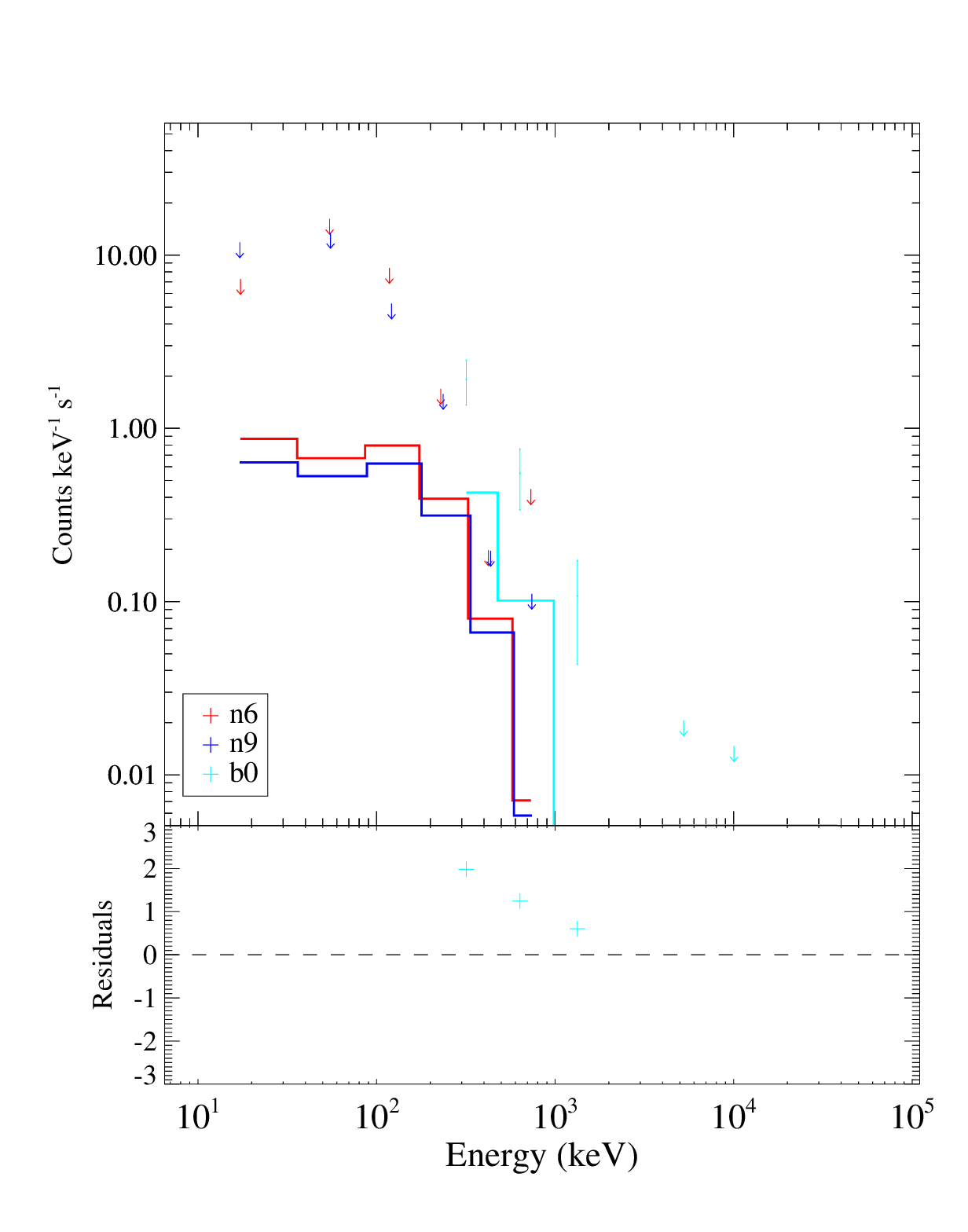}
	\includegraphics[angle=0,width=0.3\textwidth]{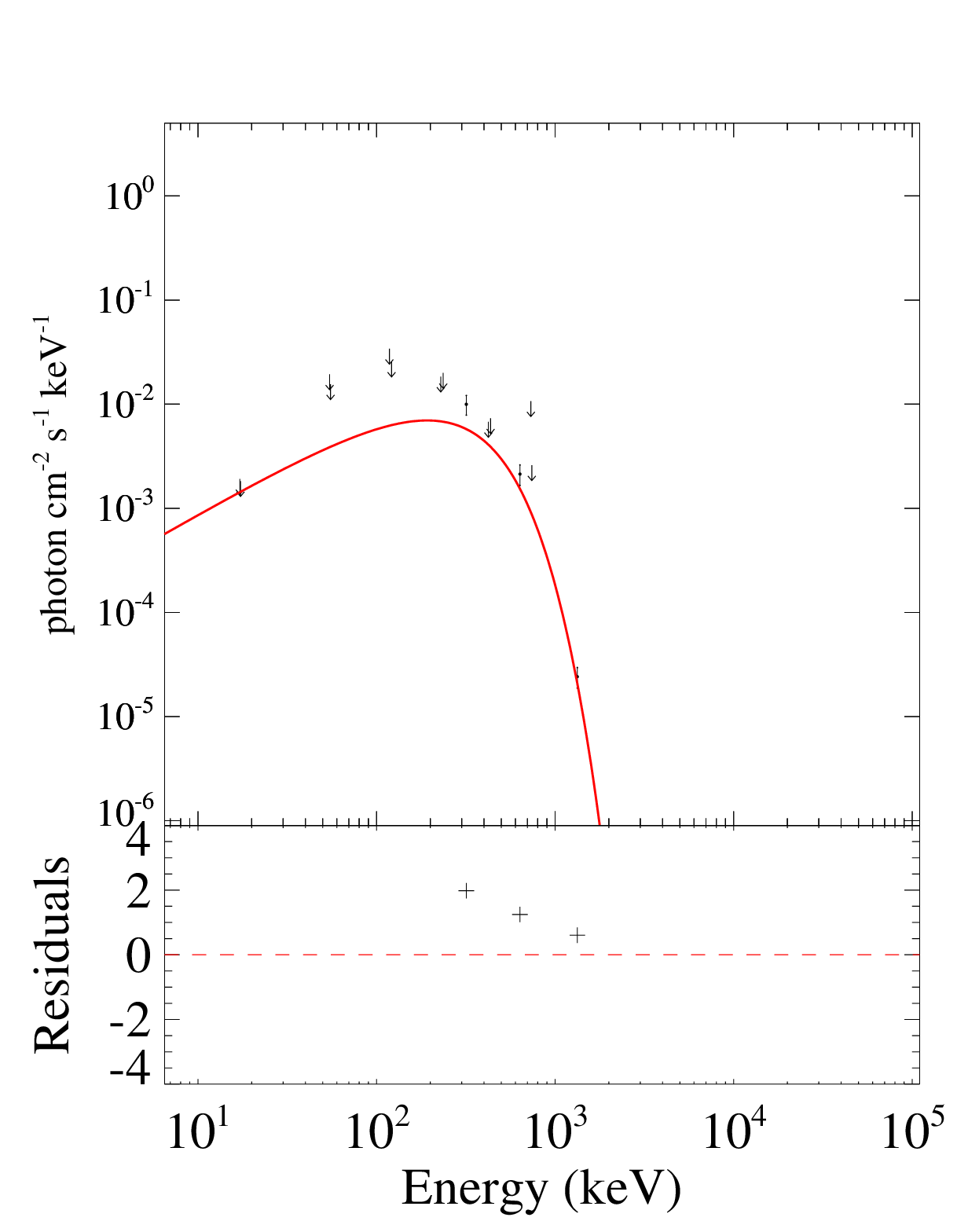}
	\includegraphics[angle=0,width=0.38\textwidth]{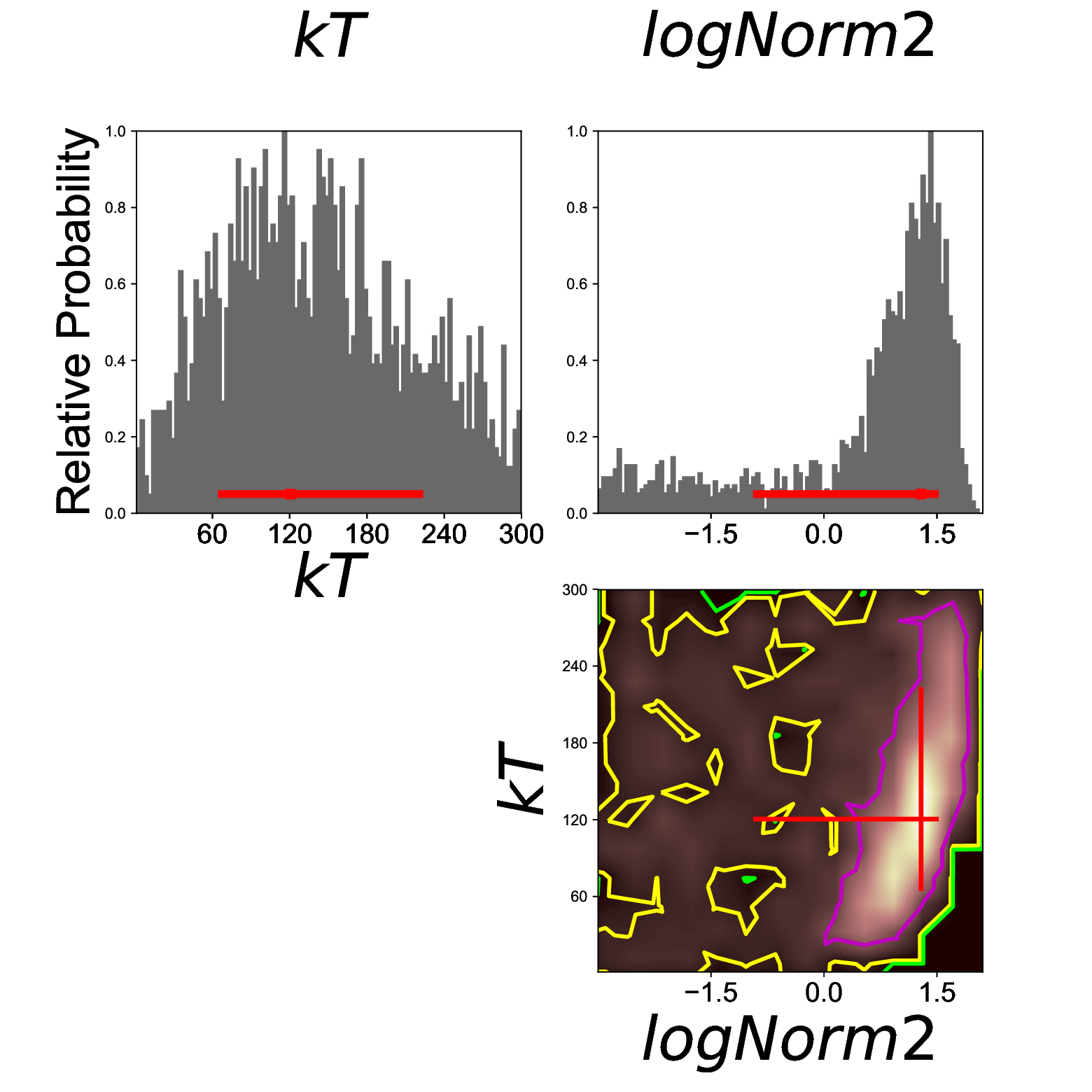}
	\caption{Constraints on the blackbody model of the precursor of GRB090510016.}\label{fig:GRB090510016}
\end{figure}

\begin{figure}[tbp]
\centering
	\includegraphics[angle=0,width=0.3\textwidth]{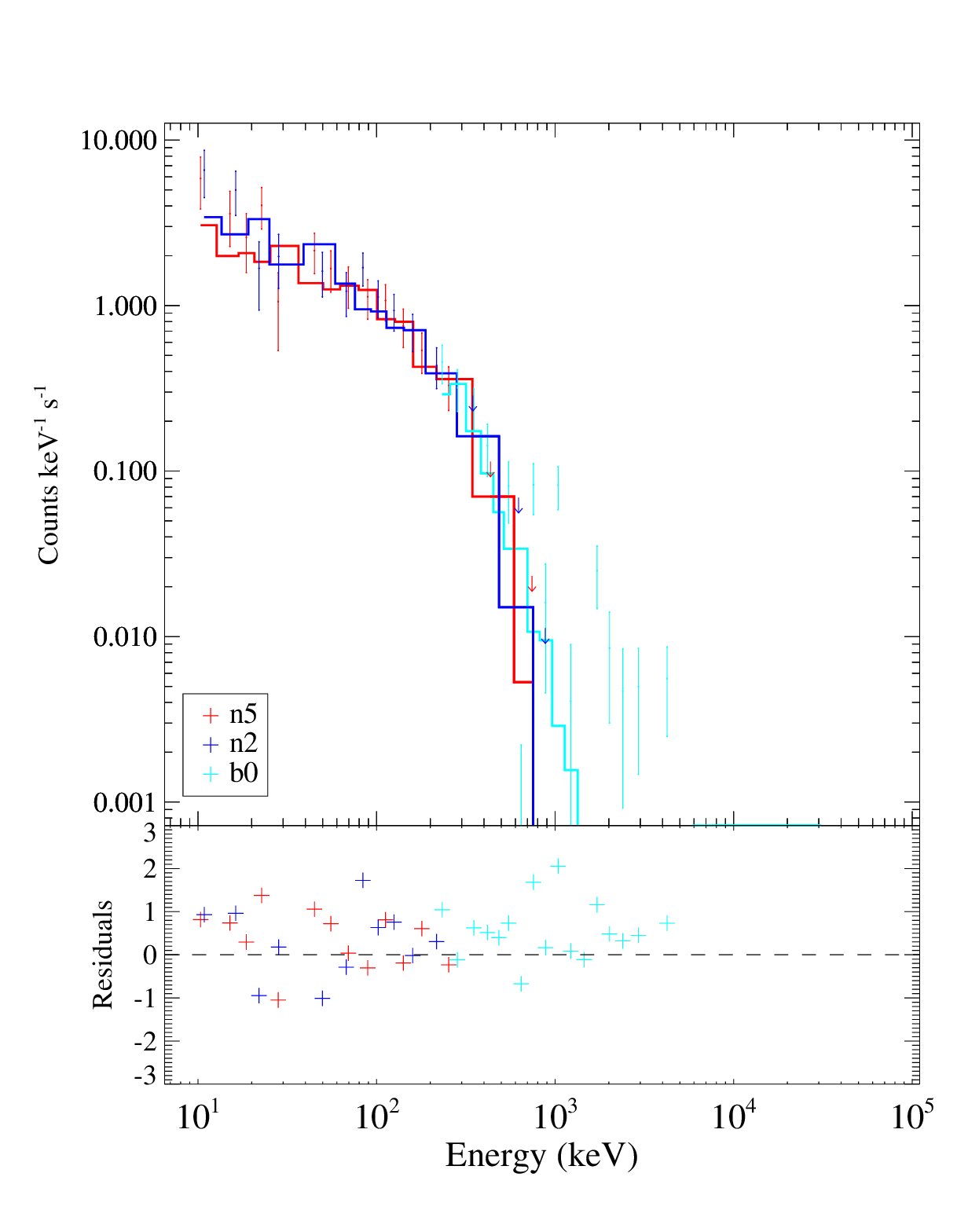}
	\includegraphics[angle=0,width=0.3\textwidth]{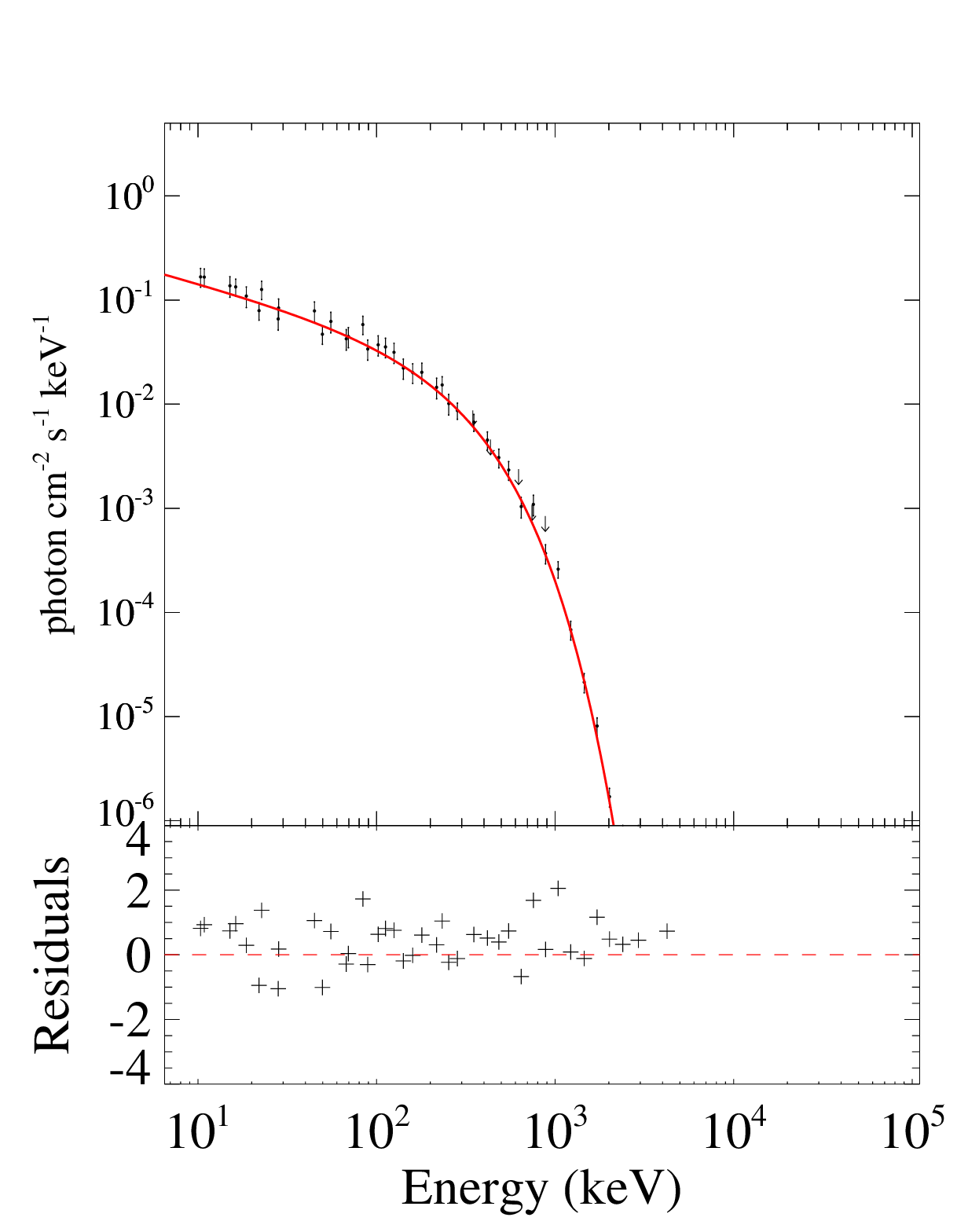}
	\includegraphics[angle=0,width=0.38\textwidth]{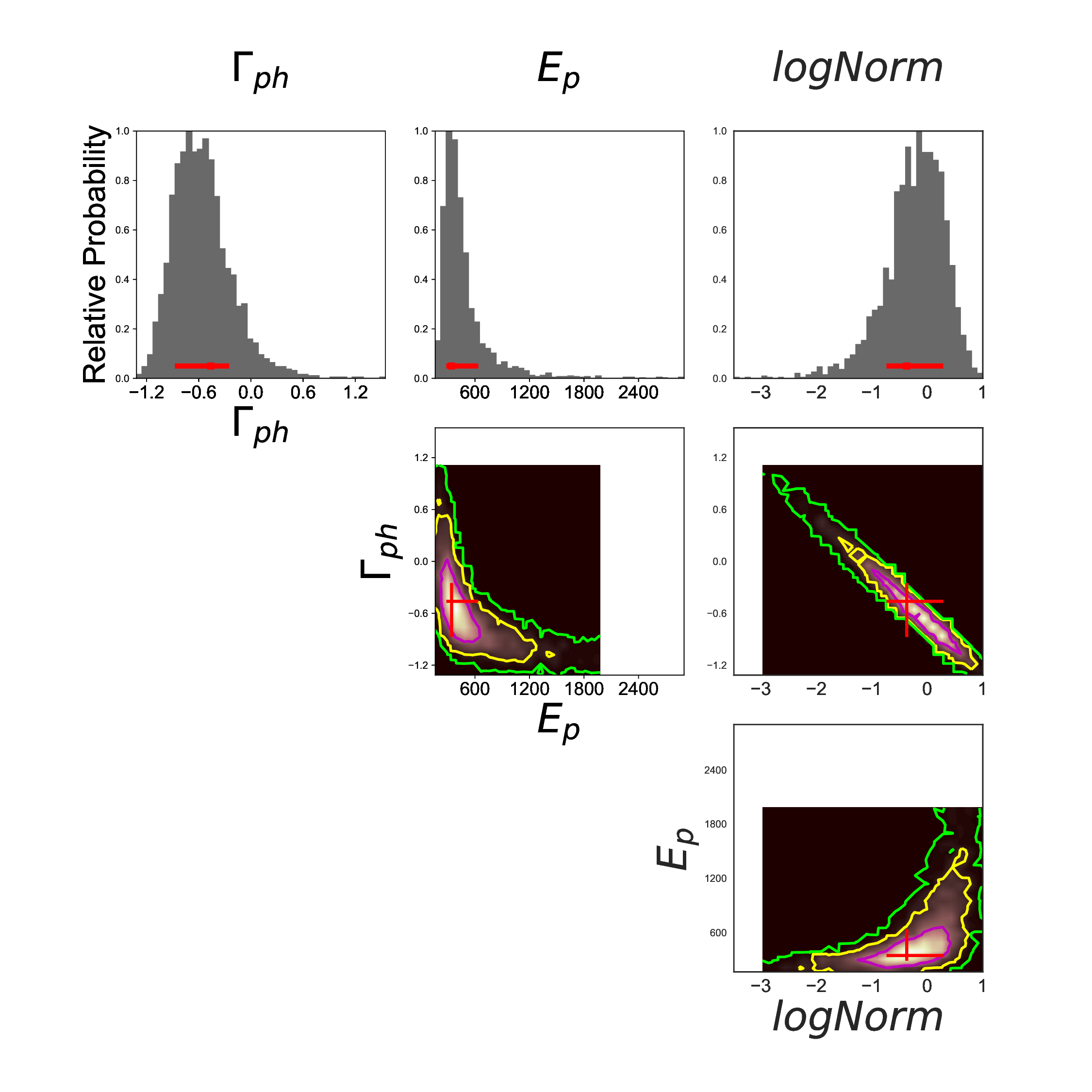}
	\caption{Constraints on the CPL model of the precursor of GRB160804180.}\label{fig:GRB160804180}
\end{figure}

Using the above three requirements, we identify 16 precursor events of SGRBs in the {\em Fermi}/GBM catalogue, accounting for 3.0\% of the full sample.
Albeit we set $3\sigma$ as the threshold, we find the significance level of our precursor sample satisfies $\gtrsim4.5\sigma$.
The light curves of these SGRBs using both the ordinary histograms and BB algorithms, as well as the evolution of the hardness ratio, are shown in Fig. \ref{fig:1}. To further study their spectral properties, we employ the McSpecFit package \citep{Zhang2018a} to perform the spectral fitting for the precursor and main SGRB emission components using the data from two nearest NaI detectors and one BGO detector. 
This package includes various spectral models, such as blackbody, BAND \citep{Band1993}, BAND+blackbody, power-law (PL), PL+blackbody, exponential cutoff power-law (CPL), and CPL+blackbody. 
The Bayesian information criterion (BIC) is used to indicate the goodness of fits to these models, where $2\leq\db<6$ gives positive evidence and $\db\geq6$ gives strong positive evidence in favor of the model with a lower BIC \citep{Robert1995}. 
Here we adopt $\db=6$ to select the best-fit model, and for those with $\db<6$, we show two favoured models.
The main features, including the duration and best fitting spectral models of the precursor and main SGRB emission components, are listed in Table \ref{tab:pre}. Fig. \ref{fig:2} shows the statistics of the durations. 
We find that in most cases, the spectral models of both the precursor and the main SGRB can be constrained, while in four cases (GRB170802638, GRB180511437, GRB181126413, and GRB191221802), only the spectral models of the main SGRB can be constrained. 
Most precursors can be fitted by the blackbody, PL or CPL models with $\db\gtrsim2$. 
Note for GRB081216531, although both blackbody and CPL models are favored with $\db=5.9$, the spectral index of CPL ($N(E)\propto E^{2.1}$) would suggest a blackbody origin. 
Three typical precursor spectra are shown in Fig. \ref{fig:GRB081216531} - \ref{fig:GRB160804180} as examples.
The best-fitting models for the main SGRBs are usually CPL or BAND models with $\db\gtrsim2$, but some can be fitted with the blackbody, PL, or CPL+blackbody models. 
Most precursors have different spectra from the main SGRBs, except GRB160804180 (both are CPL or BAND models) and GRB170709334 (both are blackbody or CPL models). 

In the top panels of Fig. \ref{fig:2}, we show the histograms of $\tp$, $\tw$, and $\tg$. In the bottom panels, we directly compare these three timescales in scatter plots. One can see that $\tw\sim\tg\sim\tp$ is generally satisfied. 
The differential number distributions of $\tp$ and $\tw$ seem to be consistent with normal distributions, but more data are required to draw a firmer conclusion. 
The precursor component has a typical duration of $\tp\lesssim0.7$ s, with a significant outlier GRB180511437 that has $\tp\approx2.8$ s. 
In most cases, the waiting time satisfies $\tw<2$ s, but there are two significant outliers: GRB180511437 with $\tw\approx13$ s and GRB191221802 with $\tw\approx19$ s. 
Using the linear regression method, we find a linear correlation in logarithmic scale, i.e. $\tw\approx2.8\tg^{1.2}$, with the correlation coefficient being $r=0.75$. However, there is also an outlier, GRB191221802 with $\tw/\tg\approx52$. 

\section{Physical Implications for Precursor Emission in SGRBs}

It has been argued that the classification of SGRBs based on $T_{90}$ could be biased for some GRBs, especially those at high redshifts (e.g., $z\gtrsim1$). These apparent SGRBs which could be intrinsically from collapsars, yet the observed light curve is just the ``tip-of-iceberg'' \cite[e.g.][]{Zhang2009,Virgili2011,Bromberg2013,Lv2014} of the emission with a longer duration. 
In our sample, the redshift of most events is unknown, except GRB090510016, which has a spectroscopic redshift $z=0.903$ \citep{Rau2009}. 
Therefore, we calculate the amplitude $f$-factor for these SGRBs \citep[see more details in][]{Lv2014} to determine the probability of some of them might be disguised SGRBs. We find that eight (four) of them have $f\gtrsim1.5(2)$, as listed in Tab. \ref{tab:pre}.  These numbers are large enough to support their NS merger origins \citep{Lv2014}. 
In the following, we mainly discuss the precursor models based on the NS merger scenario, keeping in mind that in rare cases, a collapsar origin of the SGRB cannot be ruled out.

\subsection{Precursor models}

Within the framework of NS mergers, several scenarios have been discussed in the literature that may give rise to precursor emission before the main SGRB. We discuss four possibilities below. The first two are pre-merger models and the last two are post-merger models.
\begin{itemize}
    \item {\bf The pre-merger NS crust cracking model}: For this mechanism, the dissipated energy likely is emitted in  thermal radiation, since the crust is highly optically thick. 
The energy released in this process is found to be $E_{\rm cc,46}=E_{\rm cc}/10^{46} {\rm erg} \lesssim 1$ \citep{Troja2010,Tsang2012}. 
This would heat the crust to $T_{\rm c}=E_{\rm cc}/C\lesssim 2.8\times10^8$ K, where $C\approx 10^{29}T_{\rm c}$ erg/K \citep{Yakovlev1999}. 
The corresponding luminosity from the crust surface is 
\beq
L_{\rm cc}\approx 4\pi R_*^2 a T_{\rm c}^4\lesssim 4.5\times 10^{42} E_{\rm cc,46}^2 ~{\rm erg~s}^{-1}, \label{L_cc}
\eeq 
where $a$ is the Stefan-Boltzmann constant, and the NS radius is assumed to be $R_*=10^6$ cm. 
\item {\bf The pre-merger magnetosphere interaction model}: 
The luminosity of magnetospheric interaction between two NSs can be estimated as \citep{Lai2012,Palenzuela2013,Wang2018}
\beq
L_{\rm MI}\approx 2.0\times10^{46} \eta B_{*,13}^2 
(a/ 30\;\!{\rm km})^{-7}~ {\rm erg~s}^{-1},\label{L_MI}
\eeq
where $B_*=10^{13}B_{*,13}$ G is the magnetic field of the main NS, $a$ is the separation between the two NSs, and the efficiency parameter $0.01\lesssim\eta\lesssim1$ depends on the magnetic field structure of the binary system. 
\item {\bf The post-merger shock breakout (SBO) model}: 
The SBO of the jet or cocoon from the fast component of the NS merger ejecta can release a minute fraction ($\zeta=10^{-4}\zeta_{-4}$) of the total kinetic energy of the outflow, i.e. $E_{\rm SBO}=\zeta E_{\rm iso}$ \citep{Gottlieb2018,Bromberg2018}. The luminosity of an SBO may be estimated as
\beq
L_{\rm SBO}\approx E_{\rm SBO}/t_{\rm SBO}=10^{47} \zeta_{-4} L_{\rm j,50}\tg t_{\rm SBO,-1}^{-1}~{\rm erg~s}^{-1}, \label{L_SB}
\eeq
where we used $E_{\rm iso}=L_{\rm j}\tg$, and $L_{\rm j}=10^{50} L_{\rm j,50}~{\rm erg/s}$ is the isotropic-equivalent jet luminosity. The SBO takes place at a radius of $R_{\rm SBO}\approx\Gamma^2_{\rm SBO}ct_{\rm SBO}$, where $\Gamma_{\rm SBO}$ is the Lorentz factor of the emitting region, which is $\Gamma_{\rm SBO}\sim10$ for jet breakout and $\Gamma_{\rm SBO}\sim3$ for cocoon breakout; $t_{\rm SBO}=0.1t_{\rm SBO,-1}$ s is the SBO timescale; The observed spectrum is quasi-thermal with a temperature $T_{\rm SBO}\sim \Gamma_{\rm SBO}(1-50)$ keV \citep{Gottlieb2018,Bromberg2018}.
\item {\bf The post-merger fireball photosphere model}:
The luminosity of photospheric radiation of a GRB fireball can be expressed as 
\beq
L_{\rm ph}=10^{50}\xi L_{\rm j,50}~{\rm erg/s},\label{eq:Lph}
\eeq
where $\xi={\rm min}[1,(R_{\rm c}/R_{\rm ph})^{2/3}]$ with $R_{\rm c}$ and $R_{\rm ph}$ being the coasting radius and photosphere radius, respectively \citep[see Section 7.3.3 of ][and references therein]{Zhang2018book}. 
This leads to a quasi-blackbody spectrum with a temperature  
\beq
kT_{\rm ph}=\xi kT_0= 40.9\xi_{-1}L_{\rm j,50}^{1/4}R_{0,7}^{-1/2}~{\rm keV},\label{eq:Tph}
\eeq where $T_0$ 
and $R_0=10^7R_{0,7}$ cm are the initial temperature and the size of the fireball. 

\end{itemize}

Recently, \cite{Dichiara2020} performed a systematic search for SGRBs in the local Universe based on the {\em Swift} catalog, and found that four closest SGRBs could be located at distances of 100-200 Mpc. 
The sensitivity of {\em Fermi}/GBM is roughly $0.5~{\rm cm}^{-2} {\rm s}^{-1}$ assuming a photon energy of $100$ keV\footnote{see https://fermi.gsfc.nasa.gov/science/instruments/table1-2.html}. The corresponding threshold luminosity for the events detectable at a luminosity distance of $D>100$ Mpc is
\begin{equation}
    L_{\rm th} \sim 10^{47} ~ {\rm erg \ s^{-1}} (D / 100 \ {\rm Mpc})^2.
\end{equation}
Comparing this with the predicted luminosities of the four precursor models, one can see that the crust cracking model predicts too faint precursor emission to be detectable. 
For cosmological-distance-SGRBs ($D > 100$ Mpc), only the SBO emission and fireball photosphere model can give rise to a bright enough precursor for SGRBs. The magnetosphere interaction model may be relevant to presursor emission of some SGRBs if the sources are nearby and the surface magnetic field of the primary NS is strong enough (e.g. $B_s > 10^{13}$ G).

\subsection{Constraints on GRB models}

Some precursors in our sample can be explained by the blackbody model with $\db\gtrsim2$, especially GRB081216531 and GRB141102536 with $\db\gtrsim6$ (see in Table \ref{tab:pre}). This is consistent with the SBO and fireball photosphere model. 
The observed relative flux ratio between the precursor and to the main SGRB is about $0.01\lesssim L_{\rm pre}/L_{\rm j}<1$ in our sample. For the SBO model, it requires $10\lesssim \zeta_{-4}\tg t_{\rm SBO,-1}^{-1}<10^3$. 

For the fireball photosphere model, the relative flux ratio as well as the precursor temperature can be well explained by the model with $1>\xi\gtrsim0.01$. 
The observed duration of the photospheric radiation is characterized by $t_{\rm ph}\approx R_{\rm ph}(1+z)/(\Gamma^{2}c)$, where $z$ is the redshift, $\Gamma$ is the bulk Lorentz factor of the jet, and the photosphere radius is $R_{\rm ph}=5.9\times10^{13}L_{\rm j,50}\Gamma_1^{-3}~{\rm cm}$, where $\Gamma_1=\Gamma/10$ \cite[e.g.][]{Meszaros2000,Rees2005,Zhang2018book}. 
Our sample shows that $t_{\rm ph}\approx0.1t_{\rm ph,-1}$ s, which gives an interesting constraint on the bulk Lorentz factor of SGRB outflow, i.e.
\beq
\Gamma=28.8L_{\rm j,50}^{1/5} t_{\rm ph,-1}^{-1/5}.
\eeq
This result is consistent with Eq. (1) of \cite{Troja2010}.
Note this interpretation requires a matter-dominated jet, with the main SGRB signal originating from internal shocks \citep{Meszaros2000,Zhang2018book}. 

For the post-merger precursor models, the waiting time between the precursor and the main burst corresponds to the observer-frame time for the jet to propagate from the precursor radius $R_{\rm pre}$ (photospheric radius or SBO radius) to the jet dissipation radius $R_{\rm GRB}$, i.e., $\tw=(R_{\rm GRB}-R_{\rm pre})(1+z)/(\Gamma^{2}c)$. 
Observations show $\tw\sim\tp$ (see Fig. \ref{fig:2}), which indicates that $R_{\rm GRB}\sim2R_{\rm ph}$ for the fireball photosphere model, and $R_{\rm GRB}\sim \Gamma^2 \Gamma^{-2}_{\rm SBO} R_{\rm SBO}$ for the SBO model.
However, we should keep in mind that the definition of $\tp$ and $\tg$ here are based on $T_{90}$, which could underestimate the intrinsic durations of the precursor and the main burst and over-estimate the waiting time. 
The main GRB signal is expected be non-thermal, which is consistent with our spectral fits for most events. One exception is GRB170709334, which favors thermal spectra for both the precursor and the main GRB. This may correspond to an SBO precursor with a fireball photosphere induced main pulse or two episodes of central engine activities with the internal shock emission suppressed. 

In some cases, the precursor emission has a non-thermal spectrum, especially GRB111117510 and GRB160804180 with $\db\gtrsim6$. These cases may be explained by the NS magnetospheric interaction model (assuming that the sources are nearby). 
For NS mergers with the surface magnetic field $B_{*,13}\gtrsim1$ for the primary NS, the typical spectrum may be approximately described by a synchrotron radiation spectrum of a photon index around $-2/3$ peaking at $\sim$MeV, because of the effect of synchrotron-pair cascades \citep{Wang2018PRD,Wang2018}. 
Such a model can well explain the photon indices and peak energies of the non-thermal precursor bursts, e.g., GRB111117510, GRB140209313, and GRB160804180 \citep{Wang2018}. 
The precursor emission time for this magnetospheric interaction model roughly coincides with the gravitational wave radiation chirp signal time. So the waiting time between the precursor and the main burst should correspond to the time delay between the GW signal and the SGRB signal. 
This timescale consists of three parts \citep{Zhang2019}: the time ($\Delta t_{\rm jet}$) for the jet to be launched by the central engine, the time  ($\Delta t_{\rm bo}$) for the jet to propagate through and break out from the circum-burst medium, and the time  ($\Delta t_{\rm GRB}$) for the jet to reach the energy dissipation radius (e.g., the photospheric radius or the internal shock radius). 
The last term is $\Delta t_{\rm GRB}/(1+z)\sim \tg/(1+z) \sim0.01-1$ s, while the first two terms depend on the jet launch models. 
According to the Table 1 in \cite{Zhang2019}, for most models, $(\Delta t_{\rm jet}+\Delta t_{\rm bo})/(1+z)=0.01-1$ s. 
Consequently, one would also expect $\tw\sim\tg$. 
An exception is the SMNS/SNS magnetic model, in which a uniform-rotation-supported supramassive NS (SMNS) is formed after the NS merger, which subsequently becomes a stable NS (SNS). 
In this model, the waiting time is dominated by the term $\Delta t_{\rm jet}/(1+z)=0.01-10$ s, which is mainly contributed by the time needed to clean the environment to launch a relativistic jet \citep{Metzger2011,Zhang2019}. 
In this case, one expects $\tw\gg\tg$. 
In our sample, we find most events satisfy $\tw\sim\tg$, except GRB191221802, which has $\tw/\tg\approx52$ and $\tw=19.36_{-3.19}^{+1.24}$ s. 
We also notice that for GRB090510016, \cite{Troja2010} found two precursors in the {\em Swift} data, but only the second precursor can be found in {\em Fermi} data (consistent with our results). 
Its first precursor is found to be of $\tw/\tg\approx40$ and $\tw\approx12$ s, while its second precursor in our analysis is consistent with the photospheric radiation of the fireball. 
Therefore, its first precursor with a long waiting time ($\tw/\tg\gg1$) could originate from NS magnetospheric interaction, and such long waiting times are caused by the jet launch mechanism in the SMNS/SNS magnetic model. 
In conclusion, according to this model, a SNS engine might have been formed after the merger in events with $\tw/\tg\gg1$, e.g. GRB090510016 and GRB191221802. 

\section{Conclusions and discussion}

\begin{figure}
	\centering
	\begin{subfigure}
		\centering
		\includegraphics[width=\textwidth]{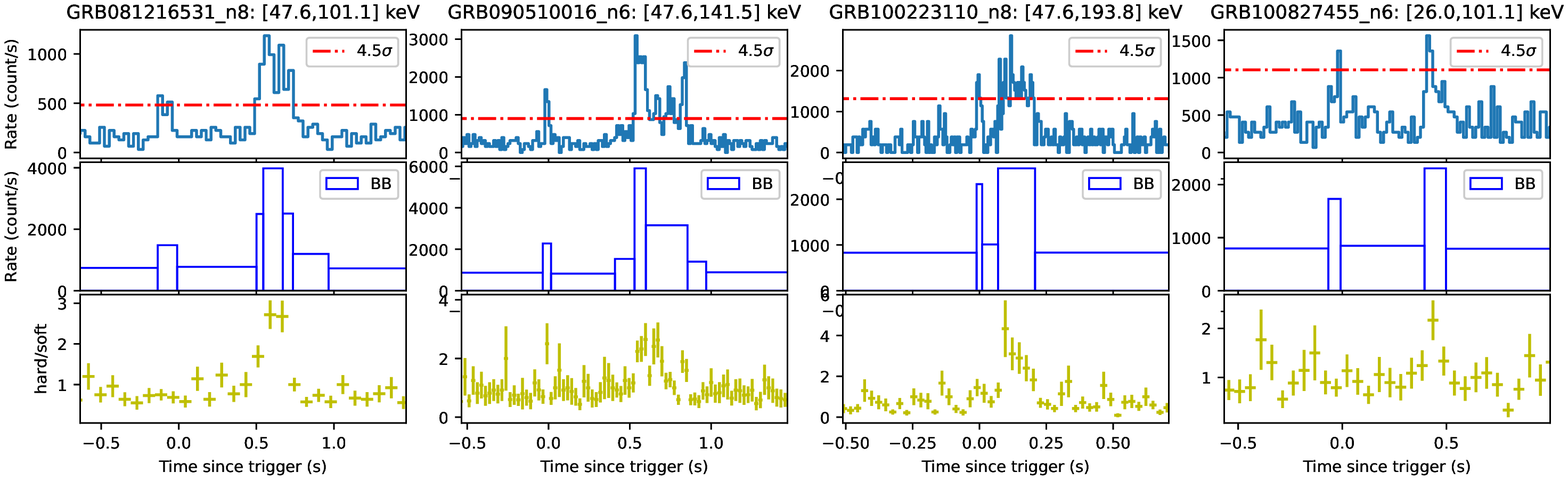}
	\end{subfigure}
	\vskip-0.3cm
	\begin{subfigure}
		\centering
		\includegraphics[width=\textwidth]{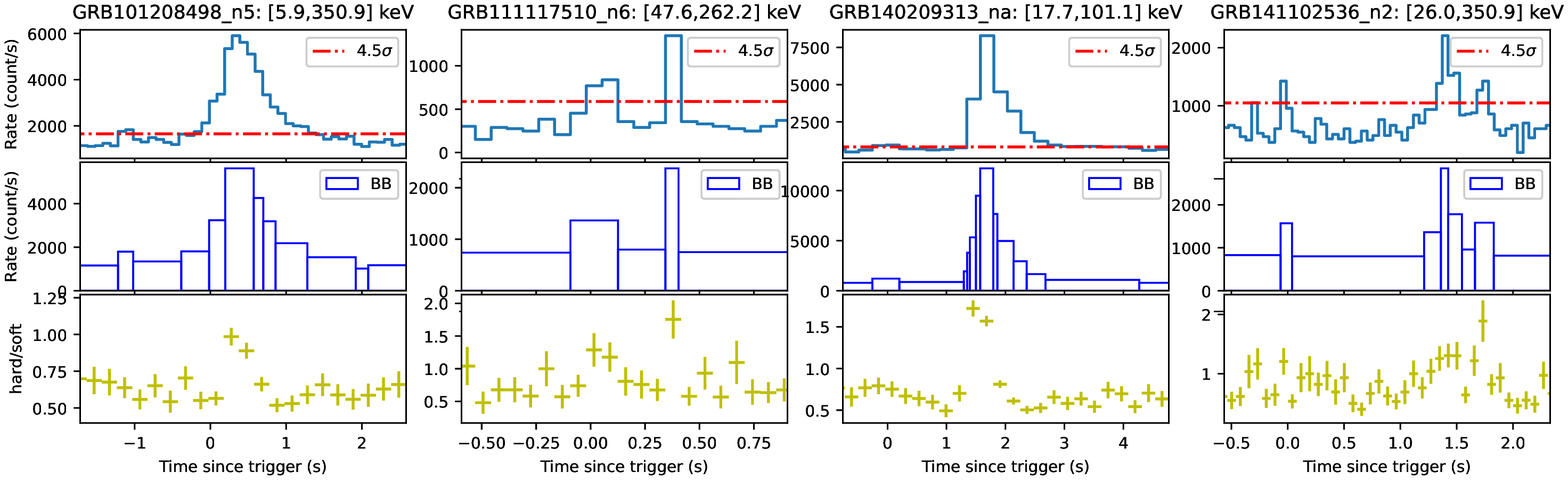}
	\end{subfigure}
	\vskip-0.3cm
	\begin{subfigure}
		\centering
		\includegraphics[width=\textwidth]{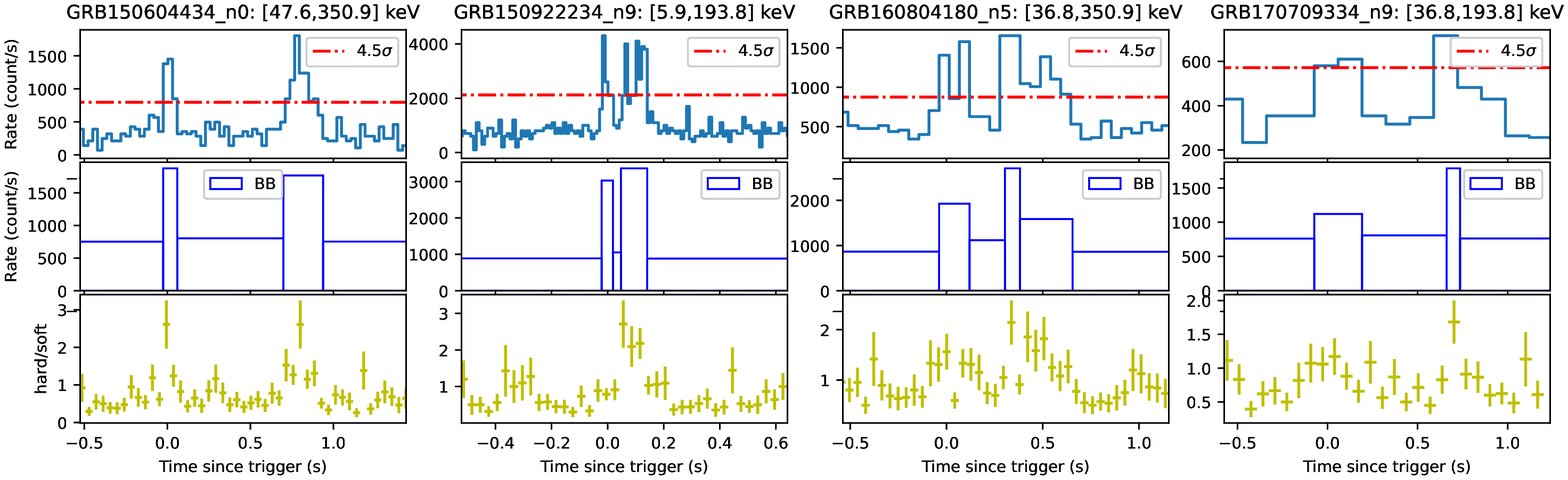}
	\end{subfigure}
	\vskip-0.3cm
	\begin{subfigure}
		\centering
		\includegraphics[width=\textwidth]{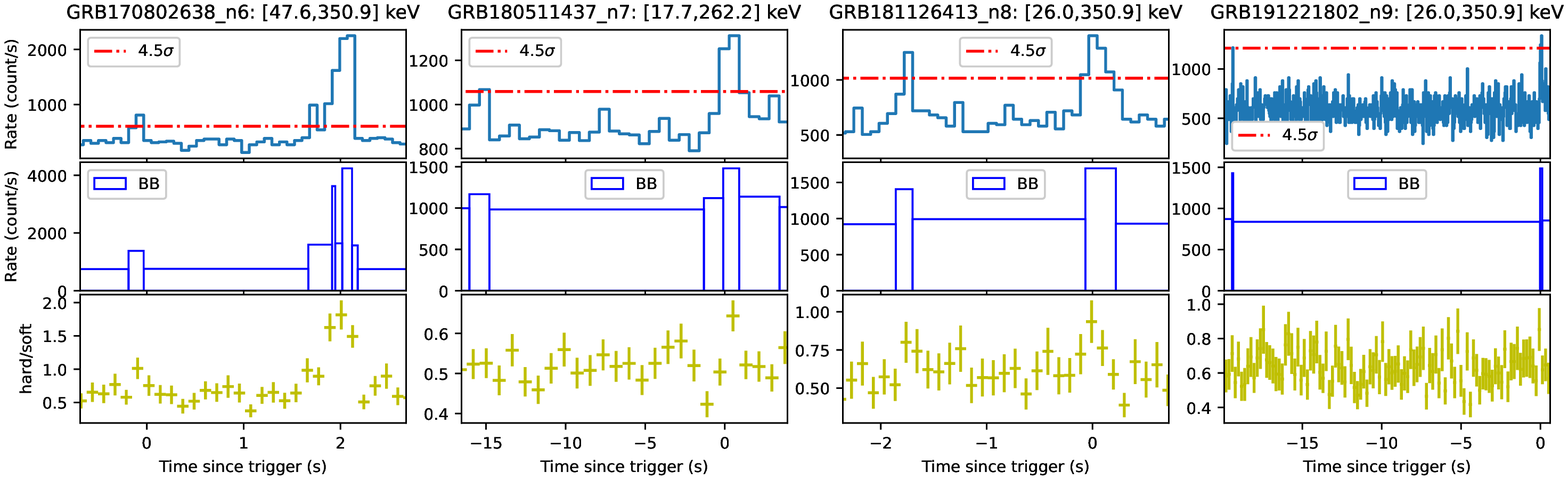}
	\end{subfigure}
	\caption{The light curves of SGRBs from the NaI detector with the highest significance for their precursors. We use both the traditional histogram and BB algorithm. The traditional histograms are obtained for the specific energy range optimized for the significance level of the precursor. Note that for GRB 150922234, the  peak flux of the precursor appears larger than that of the main pulse, but it is smaller than in the light curve of the full energy band (and hence, is defined as precursor emission). The hardness ratio (hard/soft) is the ratio of numbers between hard photons ($50-800$ keV) and soft photons ($10-50$ keV). \label{fig:1}}
\end{figure}

\begin{figure}
	\includegraphics[width=\textwidth]{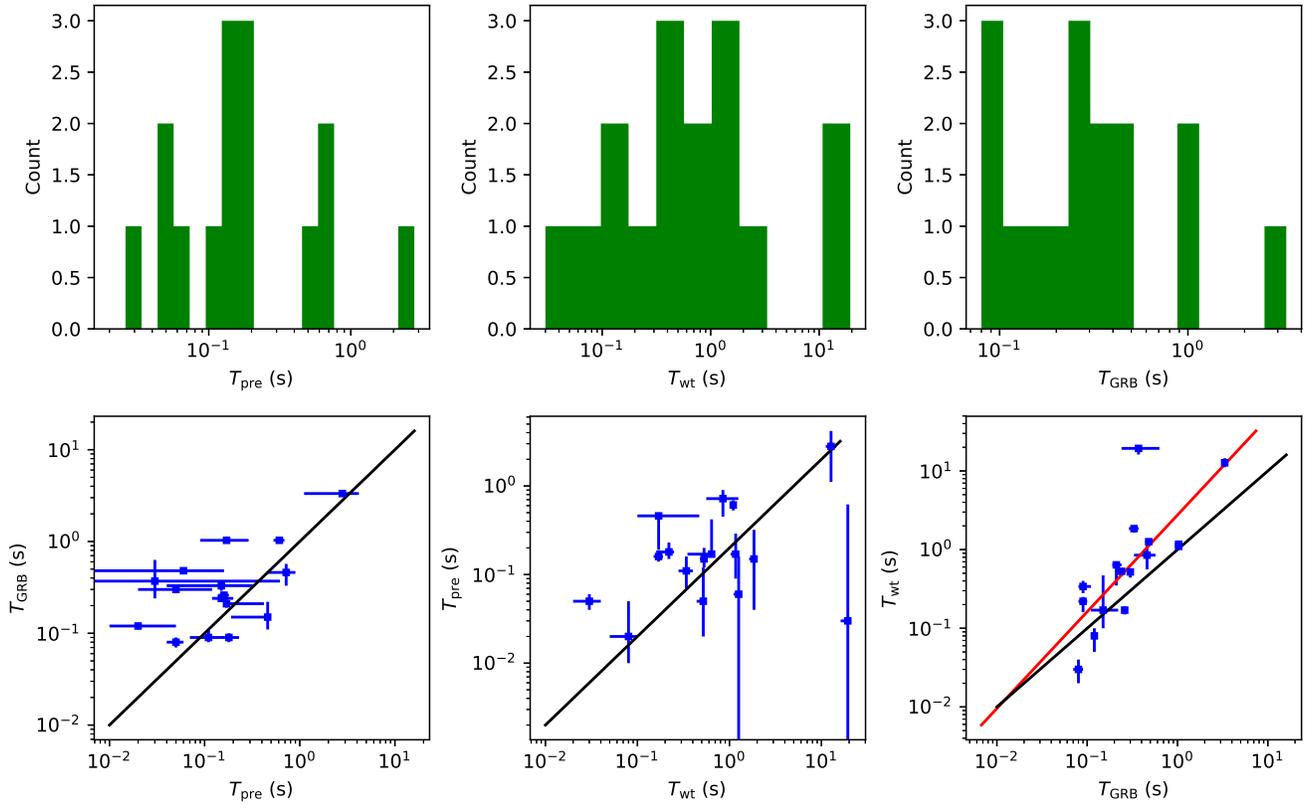}
	\caption{The top panels show the histograms of $\tp$, $\tw$, and $\tg$. 
		The bottom panels show comparisons of these timescales, and the black lines represent the equality line. 
		The red line in the right bottom panel is $\tw\approx2.8\tg^{1.2}$. \label{fig:2}}
\end{figure}

In this paper, we performed a stringent search for precursor emission of short GRBs in the {\em Fermi}/GBM data and found that 16 out of 529 (3.0\%) SGRBs have precursor with significance $\gtrsim4.5\sigma$. 
The light curves are shown in Fig. \ref{fig:1} and the properties of precursor and main SGRB emission are listed in Tab. \ref{tab:pre}. 
As shown in Fig. \ref{fig:2}, the timescales are roughly comparable to each other, $\tw\sim\tg\sim\tp$, and there is a linear correlation (correlation coefficient $r=0.75$) $\tw\approx2.8\tg^{1.2}$ in the logarithmic scale, but with a significant outlier $\tw/\tg\approx52$ in GRB191221802. 
In most cases, we find $\tp\lesssim0.7$ s and $\tw<2$ s, but there are significant outliers, i.e. $\tp\approx2.8$ s and $\tw\approx13$ s for GRB180511437, and $\tw\approx19$ s for GRB191221802. 

Most precursors favour the blackbody, CPL and/or PL spectra with $\db\gtrsim2$. In particular, GRB081216531 and GRB141102536 favours blackbody model with $\db\gtrsim6$, and GRB111117510 and GRB160804180 favours CPL model with $\db\gtrsim6$. 
The thermal spectra can be explained within the SBO model and the photospheric radiation fireball model, and the non-thermal ones may be explained in the NS magnetospheric interaction model. 
The crust cracking mechanisms generally predict too faint emission to be detected at a cosmological distance. 
For the SBO model, we constrain $10^{-2}\tg/\tp\lesssim \zeta<1$. This is larger than the expected value of $\zeta\sim10^{-4}-10^{-3}$ \citep{Gottlieb2018,Bromberg2018}. One possible explanation is that the jet is viewed slightly off-axis so that the observed luminosity of the main pulse is smaller than the jet luminosity.
For the photospheric radiation mechanism, a matter-dominated jet is preferred. We constrain the jet Lorentz factor to be $\Gamma=28.8L_{\rm j,50}^{1/5} t_{\rm ph,-1}^{-1/5}$. 
However, as noted by \cite{Troja2010}, such a Lorentz factor is much smaller than that of typical SGRBs and thus may have difficulties to explain SGRB properties. For example, observation shows that the Lorentz factor of GRB090510016 should be $\Gamma\gtrsim 10^3$ \citep{Ackermann2010}.
For the NS magnetospheric interaction model, we find it can provide a constraint on jet launch mechanism. 
More specifically, we find events with $\tw/\tg\gg1$ can be well explained by the time delay to launch a relativistic jet in the SMNS/SNS magnetic model.  
As a consequence, in GRB191221802 there might be an SNS formed after each merger, and their jets are magnetically powered. 

We also notice that the possibility that some events in our sample are from collapsars cannot be excluded. 
For example, it is unclear whether GRB180511437 is a short GRB or not, as our study shows  $\tg=3.33_{-0.24}^{+0.18}$ s, which is $>2$ s, even though it has $\tg= 1.98\pm0.97$ in the {\em Fermi} GBM Burst Catalog\footnote{see on webpage \url{https://heasarc.gsfc.nasa.gov/db-perl/W3Browse/w3query.pl}}.
Besides, the precursor of GRB160804180 can be also  explained by the BAND model with $\db=5.9$, which might be an early episode activity from the central engine.
Furthermore, it is also suggested that GRB090510016 could be of collapsar origin based on the study of its afterglow \citep{Zhang2009,Panaitescu2011}.
Such grey-zone cases can be better studied when multi-wavelength/multi-messenger information (e.g. host galaxy identifications) becomes available \citep[e.g.][]{Liye2020,Dichiara2020}. 

Albeit only 3.0\% SGRBs detected by Fermi/GBM have detectable precursor emission, we note that the opening angles of precursors from SBOs (especially for cocoon breakouts) and NS magnetospheric interactions can have a solid angle much larger than the jet opening angle. 
Therefore, searching EM counterparts of NS mergers in the local Universe will very likely  detect such precursor emissions with/without detecting the main SGRBs. GW170817/GRB 170817A may be such a case.

\begin{sidewaystable}
	\scriptsize
	\caption{ The SGRBs with precursor emission and their properties} \label{tab:pre}
		\begin{tabular}{ccccccccc}
		\hline
		Name$^a$ & $\tp$ & Best-fit models of precursors$^b$&$\db$ & $\tw$ & $\tg$ & Best-fit model of main pulse$^a$&$\db$ & $f$-factor\\
	    &   (s) &  Energy unit (keV)& & (s) & (s) & Energy unit (keV)& &\\\hline
		
		GRB081216531 & $0.15_{-0.03}^{+0.05}$ & Blackbody: $kT=18.66_{-2.11}^{+3.30}$; & $\geq5.9$ & $0.53_{-0.05}^{+0.04}$ & $0.24_{-0.02}^{+0.02}$ & CPL + Blackbody: $\Gamma_{ph}= 0.08_{-0.03}^{+0.27}$, &$\geq5.0$&$2.0\pm0.1$\\
		&  &CPL: $\Gamma_{ph}=2.1^{+1.20}_{-1.73}$, $E_{p}=72.49^{+17.17}_{-8.61}$  &  &  & & $E_{p}=265.94_{-30.44}^{+36.65}$, $kT=340.85_{-21.47}^{+31.98}$;  & &\\
		&&&&&&CPL: $\Gamma_{ph}=-0.50^{+0.06}_{-0.06}$, $E_{p}=1219.0^{+103.1}_{-114.5}$&\\\hline
		
		GRB090510016$^{*,c}$  & $0.05_{-0.03}^{+0.07}$& Blackbody: $kT=120.43_{-56.11}^{+103.18}$; &$\geq3.0$ & $0.52_{-0.08}^{+0.04}$ & $0.30_{-0.01}^{+0.01}$ & CPL: $\Gamma_{ph}=-0.61^{+0.03}_{-0.02}$, $E_{p}=2999.44^{+0.55}_{-62.77}$&$\geq79.7$&$2.6\pm1.0$\\
		&&PL: $\Gamma_{ph}=-1.13^{+1.89}_{-4.19}$&&&&&\\\hline
		
		GRB100223110  & $0.02_{-0.01}^{+0.03}$ & Blackbody: $kT=66.02_{-15.62}^{+135.35}$; &$\geq1.7$ & $0.08_{-0.03}^{+0.02}$ & $0.12_{-0.01}^{+0.01}$ & CPL: $\Gamma_{ph}=-0.18^{+0.11}_{-0.12}$, $E_{p}=1101.63_{-107.93}^{+181.12}$;&$\geq5.8$&$1.4\pm0.1$\\
		&  & PL: $\Gamma_{ph}=-1.17^{+0.09}_{-4.01}$ &  && \multicolumn{3}{c}{BAND: $\alpha=-0.19^{+0.12}_{-0.11}$,$\beta=-13.65^{+7.55}_{-3.52}$, $E_p=1122.3^{+153.4}_{-123.9}$} &\\\hline
		
		GRB100827455  & $0.11_{-0.04}^{+0.05}$ & Blackbody: $kT=98.60_{-37.80}^{+145.67}$; & $\geq4.1$& $0.34_{-0.06}^{+0.06}$ & $0.09_{-0.01}^{+0.02}$ & Blackbody: $kT=168.19_{-57.24}^{+82.07}$; &$\geq1.7$ & $1.4\pm0.3$\\
		&  & PL: $\Gamma_{ph}=-1.47^{+0.14}_{-3.74}$ &  &  & & PL: $\Gamma_{ph}=-1.11^{+0.17}_{-3.74}$;&&\\ \hline

		GRB101208498  & $0.17_{-0.08}^{+0.12}$ &  Blackbody: $kT=9.74_{-1.68}^{+1.90}$;  &$\geq1.1$& $1.17_{-0.14}^{+0.10}$ & $1.03_{-0.04}^{+0.03}$ & CPL: $\Gamma_{ph}= -0.77_{-0.07}^{+0.06}$, $E_{p}=148.24_{-6.77}^{+9.76}$;&$\geq3.8$&$3.6\pm0.2$\\
		&  &  PL: $\Gamma_{ph}=-2.20^{+0.20}_{-0.44}$; &  &  & \multicolumn{3}{c}{BAND: $\alpha=-0.67^{+0.17}_{-0.17}$, $\beta=-2.63^{+0.34}_{-14.26}$, $E_p=127.6^{+29.4}_{-15.9}$}&\\\hline
		
		GRB111117510$^*$  & $0.18_{-0.03}^{+0.05}$ & CPL: $\Gamma_{ph}= -0.47_{-0.32}^{+0.22}$, $E_{p}=576.84_{-91.69}^{+442.45}$;  &$\geq6.0$& $0.22_{-0.06}^{+0.03}$ & $0.09_{-0.01}^{+0.01}$ & Blackbody: $kT=55.31_{-6.86}^{+10.38}$; &$\geq2.1$& $1.3\pm0.1$\\
		&&&&&&CPL: $\Gamma_{ph}= -0.02_{-0.48}^{+0.70}$, $E_{p}=254.25_{-39.43}^{+104.53}$&&\\\hline
		
		GRB140209313$^{*,d}$  & $0.61_{-0.08}^{+0.08}$ &  CPL: $\Gamma_{ph}=-1.07^{+0.92}_{-0.60}$, $E_{p}=114.74_{-49.19}^{+1526.38}$;    &$\geq2.3$  & $1.10_{-0.08}^{+0.08}$ & $1.03_{-0.06}^{+0.04}$ & BAND: $\alpha=-0.31^{+0.06}_{-0.05}$,&$\geq61.9$ & $7.3\pm0.6$\\
		&  &  PL: $\Gamma_{ph}=-1.74^{+0.06}_{-0.10}$ &&  &  &$\beta=-2.44^{+0.07}_{-0.08}$, $E_p=139.66^{+6.33}_{-5.77}$;&& \\\hline
		
		GRB141102536$^*$  &  $0.06_{-0.06}^{+0.10}$ & Blackbody: $kT=83.92_{-12.07}^{+38.77}$;  &$\geq6.0$& $1.26_{-0.15}^{+0.11}$ & $0.48_{-0.04}^{+0.04}$ & CPL: $\Gamma_{ph}= -0.52^{+0.15}_{-0.15}$, $E_{p}=402.76_{-46.93}^{+88.86}$;& $\geq5.7$& $1.6\pm0.1$\\
		&&&&&\multicolumn{3}{c}{BAND: $\alpha=-0.53^{+0.14}_{-0.16}$,$\beta=-3.53^{+1.19}_{-13.6}$, $E_p=405.9^{+91.9}_{-48.8}$}& \\\hline
		
		GRB150604434  & $0.17_{-0.01}^{+0.25}$ & Blackbody: $kT=124.78_{-17.16}^{+31.32}$;  &$\geq3.1$& $0.64_{-0.29}^{+0.02}$ & $0.21_{-0.02}^{+0.03}$ & CPL: $\Gamma_{ph}= -0.35^{+0.24}_{-0.28}$, $E_p=414.84^{+198.21}_{-73.32}$;&$\geq5.4$ &$1.5\pm0.1$\\
		&&CPL: $\Gamma_{ph}= 0.04^{+0.79}_{-0.40}$, $E_p=637.59^{+260.37}_{-142.66}$&&&\multicolumn{3}{c}{BAND: $\alpha=-0.13^{+0.08}_{-0.49}$, $\beta=-2.25^{+0.31}_{-14.98}$, $E_p=293.4^{+310.1}_{-95.8}$}&\\\hline
		
		GRB150922234  & $0.05_{-0.01}^{+0.01}$ & PL: $\Gamma_{ph}= -1.91^{+0.35}_{-2.85}$;  &$\geq2.2$& $0.03_{-0.01}^{+0.01}$ & $0.08_{-0.01}^{+0.01}$ & CPL: $\Gamma_{ph}= -0.23^{+0.21}_{-0.17}$, $E_p=474.00^{+86.67}_{-64.00}$;&$\geq4.8$ &$1.3\pm0.1$\\
		&&Blackbody: $kT=7.65_{-5.99}^{+212.99}$&&&&CPL + Blackbody: $\Gamma_{ph}= 0.37_{-0.30}^{+1.13}$, &\\
		&&&&&& $E_p=651.1^{+85.2}_{-429.2}$, $kT=39.63_{-37.5}^{+463.1}$ & \\\hline
		
		GRB160804180  & $0.16_{-0.02}^{+0.02}$  & CPL: $\Gamma_{ph}=-0.46^{+0.21}_{-0.41}$, $E_p=343.30^{+292.95}_{-58.10}$;  &$\geq5.9$& $0.17_{-0.02}^{+0.02}$ & $0.26_{-0.02}^{+0.02}$ & CPL: $\Gamma_{ph}=-0.24^{+0.17}_{-0.19}$, $E_p=619.80^{+163.11}_{-77.88}$;&$\geq5.9$&$1.6\pm0.1 $\\
		&\multicolumn{3}{c}{BAND: $\alpha=-0.54^{+0.31}_{-0.35}$, $\beta=-18.57^{+13.64}_{-1.41}$, $E_p=359.4^{+312.0}_{-78.5}$}&&\multicolumn{3}{c}{BAND: $\alpha=-0.23^{+0.16}_{-0.20}$, $\beta=-19.55^{+13.61}_{-0.45}$, $E_p=623.6^{+153.9}_{-83.5}$}&\\\hline
		
		GRB170709334  & $0.46_{-0.27}^{+0.01}$ & Blackbody: $kT=62.44_{-7.05}^{+23.18}$;   &$\geq5.1$& $0.17_{-0.07}^{+0.30}$ & $0.15_{-0.04}^{+0.07}$ &Blackbody: $kT=88.49_{-10.64}^{+16.88}$;&$\geq4.4$&$1.3\pm0.1$ \\&&CPL: $\Gamma_{ph}=-0.52^{+1.92}_{-0.23}$, $E_p=723.04^{+453.48}_{-460.33}$&&&&CPL: $\Gamma_{ph}=0.63^{+0.82}_{-0.58}$, $E_p=380.01^{+119.84}_{-54.90}$ &&\\\hline
		
		GRB170802638  & $0.15_{-0.11}^{+0.17}$ & \multicolumn{2}{c}{unconstrained} 
		& $1.85_{-0.21}^{+0.14}$ & $0.33_{-0.04}^{+0.04}$ & CPL: $\Gamma_{ph}=-0.62^{+0.07}_{-0.09}$, $E_p=799.50^{+155.17}_{-85.57}$;&$\geq5.5$&$1.5\pm0.1$\\
		&&\multicolumn{2}{c}{}&&&CPL + Blackbody: $\Gamma_{ph}= 0.01_{-0.01}^{+0.17}$, &\\
		&&\multicolumn{2}{c}{}&&& $E_p=269.3^{+24.9}_{-36.5}$, $kT=339.0_{-51.9}^{+65.6}$ &\\\hline
		
		GRB180511437  & $2.80_{-1.69}^{+1.38}$& \multicolumn{2}{c}{unconstrained}  & $12.72_{-1.57}^{+1.80}$ & $3.33_{-0.24}^{+0.18}$ &CPL: $\Gamma_{ph}=-0.81^{+0.22}_{-0.27}$, $E_p=119.70^{+31.63}_{-15.83}$; &$\geq5.5$&$1.4\pm0.1$\\	
		&&\multicolumn{2}{c}{}&&\multicolumn{3}{c}{BAND: $\alpha=-0.413^{+0.41}_{-0.65}$, $\beta=-2.68^{+0.51}_{-14.44}$, $E_p=87.1^{+60.7}_{-13.4}$}&\\\hline
		
		GRB181126413$^*$  & $0.72_{-0.27}^{+0.18}$ & \multicolumn{2}{c}{unconstrained}& $0.85_{-0.29}^{+0.40}$ & $0.46_{-0.13}^{+0.11}$ & Blackbody: $kT=24.52_{-2.04}^{+3.16}$;&$\geq6.2$&$1.2\pm0.1$\\	\hline	
		
		GRB191221802 & $0.03_{-0.03}^{+0.59}$&\multicolumn{2}{c}{unconstrained}& $19.36_{-3.19}^{+1.24}$ & $0.37_{-0.13}^{+0.26}$& Blackbody: $kT=67.21_{-9.46}^{+22.62}$;&$\geq1.0$ &$1.1\pm0.1$\\
		&&\multicolumn{2}{c}{} &  &  & CPL: $\Gamma_{ph}=-0.57^{+0.48}_{-0.53}$, $E_p=471.92^{+945.05}_{-126.29}$; &\\\hline
		
	\end{tabular}
	{\\The durations of the precursor ($\tp$), waiting time ($\tw$), and the main SGRB ($\tg$) are based on $T_{90}$ analyses. The best-fit models are obtained with the BIC.\\
		$^a$ The GRBs marked with `$^*$' also triggered {\em Swift}, and can be found at \url{https://swift.gsfc.nasa.gov/archive/grb_table/}. \\
		$^b$ For the blackbody model, $k$ and $T$ are the Boltzmann constant and temperature, respectively. 
		PL ($N(E)\propto E^{\Gamma_{ph}}$) and CPL ($N(E)\propto E^{\Gamma_{ph}}\exp{[-E(2+\Gamma_{ph})/E_p]}$) represent power-law and cutoff power-law models with photon indices $\Gamma_{ph}$, and $E_{p}$ is the peak energy for the CPL model. For those unconstrained events, we find that both blackbody and PL models are favored, but there are too few photons to provide a robust constraint on the parameters. The $\db$ between the best-fit model and other models are also presented. And for $\db<6$, two favoured models are provided.\\
		$^c$ The redshift of GRB090510016 is 0.903 \citep{Rau2009}.  \cite{Troja2010} found there are two precursors in this burst from the {\em Swift} data. \\
		$^d$ From the {\em Swift} observation, GRB140209313 was found to be a SGRB with extended emission, which has durations of $T_{90}=21.25 \pm 7.98$ s and $T_{50}=0.61 \pm 0.07$ s, respectively \citep{Palmer2014}. 
	}	
\end{sidewaystable}

\acknowledgements
We thank the referee for valuable comments. 
J.S.W is supported by China Postdoctoral Science Foundation (Grant 2018M642000, 2019T120335).
B.B.Z. acknowledges the supported by the Fundamental Research Funds for the Central Universities (14380035). This work is also supported by National Key Research and Development Programs of China (2018YFA0404204) and The National Natural Science Foundation of China (Grant Nos. 11833003).


\software{McSpecFit;~Astropy}
\bibliographystyle{aasjournal}
\bibliography{ref}

\end{document}